\documentclass[conference]{IEEEtran}
\IEEEoverridecommandlockouts

\usepackage{cite}

\usepackage{amssymb,amsfonts,amsthm}
\usepackage[fleqn]{amsmath}

\usepackage{algorithmic}
\usepackage{graphicx}
\usepackage{textcomp}
\usepackage{xcolor}

\usepackage{algorithm}

\usepackage{graphicx}
\usepackage{caption}
\usepackage{subcaption}
\usepackage{url}

\usepackage{balance}

\usepackage{mdframed}

\def\BibTeX{{\rm B\kern-.05em{\sc i\kern-.025em b}\kern-.08em
    T\kern-.1667em\lower.7ex\hbox{E}\kern-.125emX}}

\usepackage{todonotes}

\DeclareCaptionLabelFormat{andtable}{#1~#2  \&  \tablename~\thetable}

\begin{document}

\title{Breaking the borders: an investigation of cross-ecosystem software packages}

\author{\IEEEauthorblockN{Eleni Constantinou}
\IEEEauthorblockA{\textit{University of Mons} \\
Mons, Belgium \\
eleni.constantinou@umons.ac.be}
\and
\IEEEauthorblockN{Alexandre Decan}
\IEEEauthorblockA{\textit{University of Mons} \\
Mons, Belgium \\
alexandre.decan@umons.ac.be}
\and
\IEEEauthorblockN{Tom Mens}
\IEEEauthorblockA{\textit{University of Mons} \\
Mons, Belgium \\
tom.mens@umons.ac.be}
}

\maketitle

\begin{abstract}
Software ecosystems are collections of projects that are developed and evolve together in the same environment. 
Existing literature investigates software ecosystems as isolated entities whose boundaries do not overlap and assumes they are self-contained. However, a number of software projects are distributed in more than one ecosystem. 
As different aspects, e.g., success, security vulnerabilities, bugs, etc., of such cross-ecosystem packages can affect multiple ecosystems, we investigate the presence and characteristics of these cross-ecosystem packages in 12 large software distributions. 
We found a small number of packages distributed in multiple packaging ecosystems and that such packages are usually distributed in two ecosystems. 
These packages tend to better support with new releases certain ecosystems, while their evolution can impact a multitude of packages in other ecosystems. Finally, such packages appear to be popular with large developer communities.
\end{abstract}



\section{Introduction}
\label{sec:intro}
Software ecosystems are formed by projects that are developed and evolve together in the same environment~\cite{Lungu2008}. 
Examples of software ecosystems include package distributions for programming languages, e.g., CRAN for R, npm for JavaScript.  
Over the last decade, the research community has been investigating the evolution of software distributions by considering both social (e.g., developer retention and turnover)~\cite{Constantinou:2016:Workshop,Constantinou:2017:ISSE,Lin2017} and/or technical aspects (e.g., the effect of social changes on project evolution, dependencies and their evolution, technical lag, sustainability)~\cite{SANER2017-Constantinou,Decan2018,Zerouali2018,Decan2018ICSME,Valiev2018FSE,German2013,Wittern:2016,Decan:2018:MSR,Decan17SANER}.


The  intensive research efforts to study ecosystem evolution aim to study and address the implications that arise from problems during the evolution of packages, and more concretely how these issues propagate to the entire ecosystem due to dependencies among packages. However, there are cases where packages are distributed in more than one ecosystem. For example, the \textit{lodash} package\footnote{\url{https://github.com/lodash/lodash}}, a utility library delivering modularity, consistency and performance, is distributed in npm\footnote{\url{https://www.npmjs.com/package/lodash}}, Clojars\footnote{\url{https://clojars.org/cljsjs/lodash}}, Maven\footnote{\url{https://mvnrepository.com/artifact/org.webjars.npm/lodash}} and NuGet\footnote{\url{https://www.nuget.org/packages/lodash/}}. So, if \textit{lodash} faces any health problems or risks of discontinuation, then all four ecosystems will be affected. The impact of such problems may differ across ecosystems depending on the packages depending on the package, either directly or transitively, within each specific ecosystem. It is therefore important to go beyond the boundaries of individual ecosystems and investigate the presence, characteristics and evolution of packages that span across multiple ecosystems.

This paper focuses on the presence and characteristics of cross-ecosystem packages. To this end, we address the following research questions:
\begin{itemize}
\item $RQ_1$ How prevalent are cross-ecosystem packages and in how many ecosystems do they reside?
\item $RQ_2$ Do cross-ecosystem packages release with the same frequency in each ecosystem?
\item $RQ_3$ To what extent are cross-ecosystem packages used by other packages in each ecosystem dependency network?
\item $RQ_4$ Do cross-ecosystem packages have different characteristics than other packages in the ecosystem?
\end{itemize}

By answering these questions, we aim to get a better understanding of the attributes and differences, if any, of cross-ecosystem packages compared to other packages in each ecosystem.

\section{Dataset and Methodology}
\label{sec:methodology}

Our dataset relies on the 2018-03-13 dump of the open source discovery service libraries.io~\cite{librariesIO}. It contains historical data of packages and their characteristics for 36 software ecosystems. 
We excluded software ecosystems that were too domain-specific, targeting specific software frameworks (e.g., Meteor), software components (e.g., WordPress and Atom), as well as those that host a subset of packages available through another already considered ecosystem (e.g., Bower manages a subset of npm), to avoid providing misleading cross-ecosystem analyses.
We selected 12 ecosystems, namely npm, Pypi, Maven, NuGet, Cargo, CocoaPods, Packagist, Rubygems, Clojars, CRAN, Hackage and CPAN totalizing 1,556,300 packages. 

For each package in the dataset, we extracted its package information (name, package manager, etc.), versions and dependencies. We also extracted the information related to the associated git repository, such as number of stars, forks, issues, etc.
Since our analyses relies on git characteristics (cross-ecosystem package identification and characteristics of cross-ecosystem versus the other packages in each ecosystem), we excluded packages without a git repository.


We identified cross-ecosystem packages by looking at their associated git repository: if (at least) two packages on two distinct ecosystems share the same git repository, then they are labelled as cross-ecosystem packages and form an x-set. 
This approach is more reliable than looking for packages with identical names, which would risk missing cross-ecosystem packages due to different naming conventions for the different ecosystems (e.g., name-py, name-js) and could lead to false positives (e.g., the development of package \textsf{json} on Cargo is unrelated to the development of \textsf{json} on npm). 

\section{Results}
\label{sec:results}
\subsection*{$RQ_1$ How prevalent are cross-ecosystem packages and in how many ecosystems do they reside?}

The first research question aims to gain an initial understanding of the prevalence and spread of cross-ecosystem packages. First, we measure the number of distinct packages corresponding to the x-sets that appear in multiple ecosystems. In the case that multiple packages in one ecosystem participate in the same x-set, we include all such packages in our analyses, thus \textit{overestimating} the number of cross-ecosystem packages. 
For example, if one x-set consists of 5 packages in npm and 3 packages in Maven, all pointing to the same repository, then all 8 packages are included in our analysis.

Overall, we found 15,389 packages that are identified as cross-ecosystem packages. They represent a total of 2,928 x-sets. 
Per ecosystem, the number of cross-ecosystem packages ranges from 3 in CPAN (0.001\% of the CPAN packages) to 9,313 in NPM (1.33\% of the NPM packages). These numbers are reported in Table~\ref{tab:rq1cross}. In all cases, these \textit{cross-ecosystem packages represent a very small fraction of all packages} that are distributed in each ecosystem.

\begin{table}[t]
  \centering
    \scalebox{1}{%
      \begin{tabular}[b]{cr|cr}\hline
        \textbf{ecosystem} & \textbf{cross/total packages} & \textbf{ecosystem} & \textbf{cross/total packages} \\ \hline
        CPAN & 3 / 35,132 & Maven &  2,656 / 135,481 \\
        CRAN & 13 / 23,831 & npm & 9,313 / 699,238 \\
        Cargo & 85 / 14,495 & NuGet & 1,253 / 118,623 \\
        Clojars & 426 / 17,104 & Packagist & 485 / 198,796 \\
        CocoaPods & 161 / 42,653 & Pypi & 442 / 125,732 \\
        Hackage & 14 / 12,385 & Rybygems & 538 / 143,825 \\ \hline
      \end{tabular}
      }
       \caption{Cross-ecosystem package statistics}
      
    \label{tab:rq1cross}%
  \end{table}%

\begin{figure}[t]
\centerline{\includegraphics[width=0.35\textwidth]{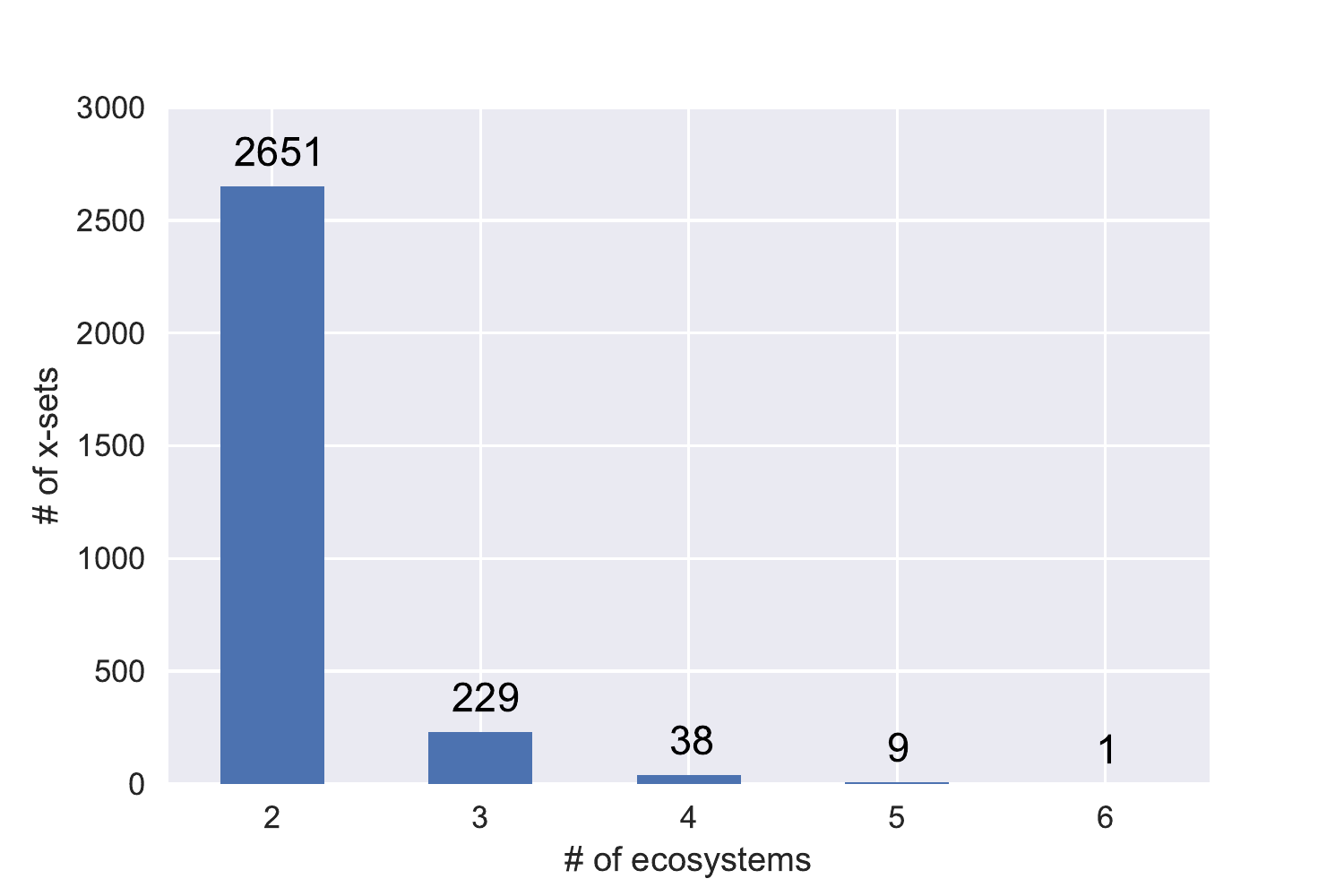}}
\caption{Number of x-sets whose packages are distributed on 2 to 6 ecosystems.}
\label{fig:cross-eco}
\end{figure}

For each x-set, we identified in how many and which ecosystems its cross-ecosystem packages are distributed. Figure~\ref{fig:cross-eco} shows the number of x-sets whose packages are distributed on 2 to 6 ecosystems. 
We observe that \textit{most x-sets spread over only 2 ecosystems}. However, we found 276 x-sets that spread over 3 to 5 ecosystems. We also found one x-set, containing the FontAwesome\footnote{\url{https://github.com/FortAwesome/Font-Awesome}} package, that spreads over 6 ecosystems. 

\begin{figure}[t]
\centerline{\includegraphics[scale=0.33]{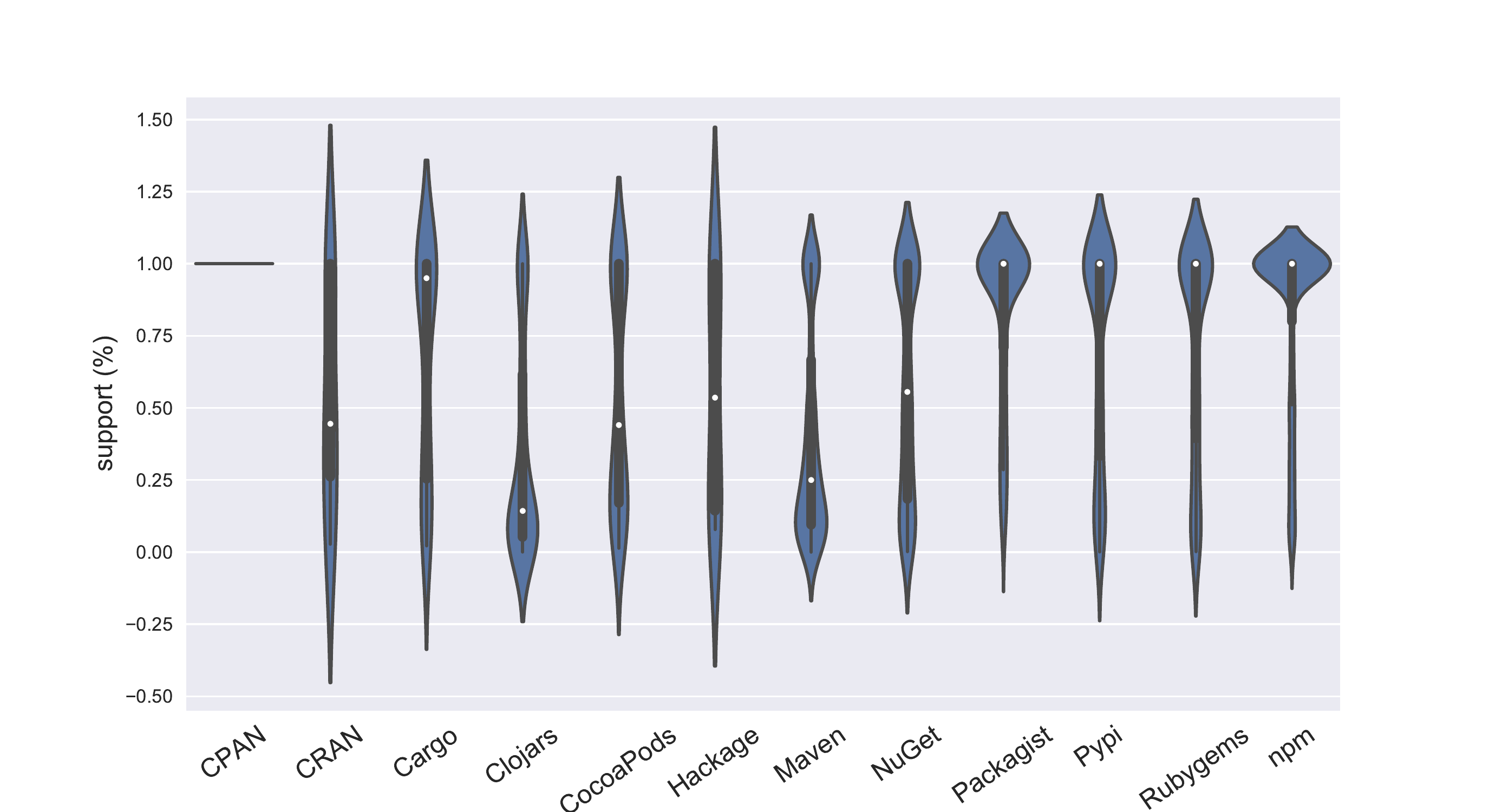}}
\caption{Distribution of $\mathrm{support}_{e}(X)$ for all x-sets $X$ and all ecosystem $e$ where $X$ is distributed.}
\label{fig:support}
\end{figure}


\subsection*{$RQ_2$ Do cross-ecosystem packages release with the same frequency in each ecosystem?}

The second research question focuses on (cross-ecosystem) package release frequency. 
For each x-set $X$, we identified in which ecosystem $e_1$ package $p\in X$ has the highest number $n$ of releases. Then, for each other ecosystem $e_2$, we computed the ratio between the number of releases that $p'\in X$ has in $e_2$ and $n$. We call this ratio the \emph{support} for $X$ in $e_2$. Formally, assuming $\mathrm{\#releases}_{e}(X)$ corresponds to the number of releases that package $p\in X$ has in ecosystem $e\in E$, 

$$
\mathrm{support}_{e_2}(X) = \frac{\mathrm{\#releases}_{e_2}(X)}{\max_{e_1\in E}{\mathrm{\#releases}_{e_1}(X)}}
$$


With this definition, at least one ecosystem has a support equal to 1, while for the remaining ecosystems in $E$ support ranges between [0, 1] (0 corresponding to an x-set whose packages are not distributed in the corresponding ecosystem). 

For each x-set $X$, and each considered ecosystem $e$ that distributes a package in $X$, we computed $\mathrm{support}_{e}(X)$. 
Figure~\ref{fig:support} shows the distribution of $\mathrm{support}_{e}(X)$ by means of violin plots. 
We observe different distributions, depending on the ecosystem. For instance, some ecosystems such as npm, Packagist, PyPI and Rubygems mostly have $\mathrm{support}$ close to the maximal value, where other ecosystems such as Clojars and Maven mostly have $\mathrm{support}$ close to the minimal value. Finally, some ecosystems stand in between, e.g., CRAN, CocoaPods, Hackage or NuGet.

These results show that \textit{maintainers tend to better support with new releases certain ecosystems}, while providing less support for others. Interestingly, we found 109 x-sets (out of 2,928) that have maximal support (i.e., $\mathrm{support} = 1$) for all the ecosystems their packages are distributed in.

\subsection*{$RQ_3$ To what extent are cross-ecosystem packages used by other packages in each ecosystem dependency network?}


\begin{figure}[t]
\centering
    \begin{subfigure}[t]{0.4\textwidth}
        \centering
        \includegraphics[scale=0.45]{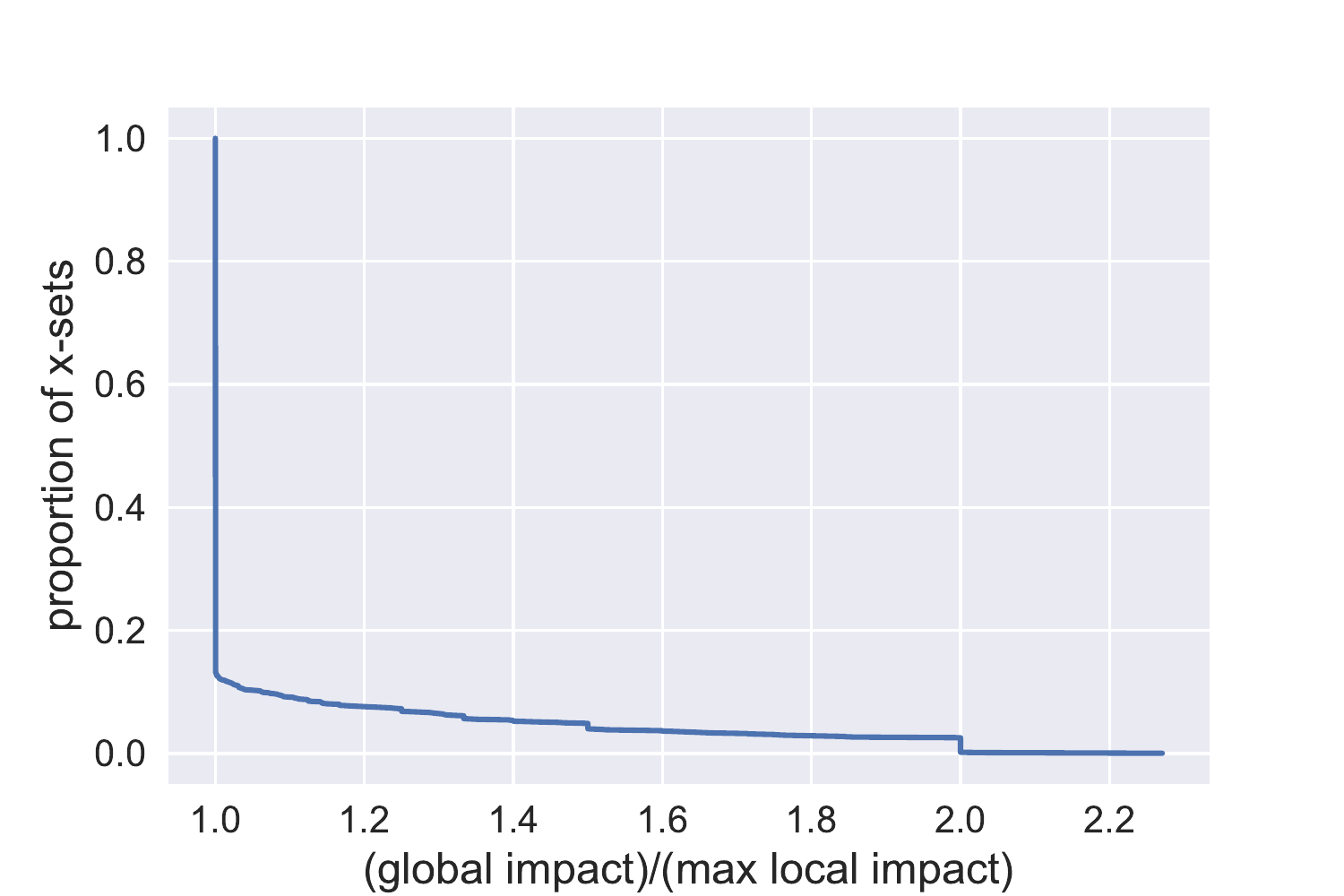}
        \caption{Direct}
        \label{fig:rq3direct}
    \end{subfigure}%
    \qquad
    \begin{subfigure}[t]{0.4\textwidth}
        \centering
        \includegraphics[scale=0.45]{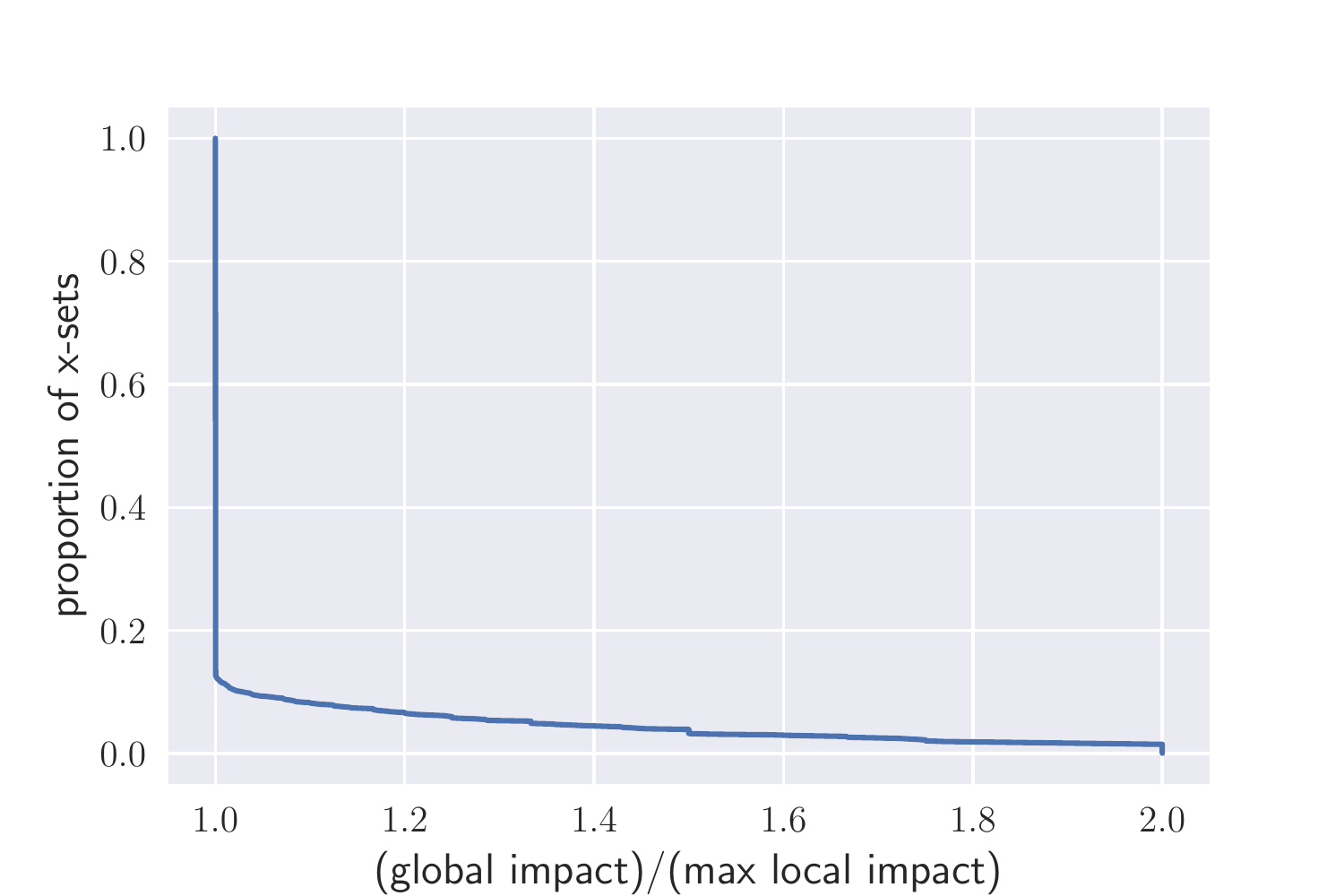}
        \caption{Transitive}
        \label{fig:rq3transitive}
    \end{subfigure}
    \caption{Reversed cumulative proportion of x-sets in function of their impact ratio.}
\end{figure}

One of the main reasons why packages depend on others is to enable software reuse. Packaging ecosystems make it easier for developers to reuse code from other packages through dependencies. On the other hand, dependencies increase the risk of having important security or maintainability problems and failures~\cite{Decan2018}. As such issues in cross-ecosystem packages could affect multiple ecosystems simultaneously, we investigate the extent that cross-ecosystem packages are used by other packages in each ecosystem. 

We limited our analysis to 6 ecosystems (npm, NuGet, Cargo, Packagist, Rubygems and CRAN), because the dependency information for these ecosystems is known to be complete and accurate, while it is not necessarily the case for the other ecosystems. The total number of  cross-ecosystem packages in these 6 ecosystems is 11,608, corresponding to 2,773 x-sets.
For each package in these x-sets, we count the number of direct and transitive dependents (i.e., the number of \emph{other} packages having a direct or transitive dependency on them) in each ecosystem the package is distributed. This number somehow represents the ``potential impact'' of a package. As cross-ecosystem packages could affect dependent packages in more than one ecosystem, we compared its ``global potential impact'' (total number of dependents across all ecosystems) and its ``maximal local impact'' (maximum number of dependents within a single ecosystem). 


Figures~\ref{fig:rq3direct} and \ref{fig:rq3transitive} show the reversed cumulative proportion of x-sets in function of their impact ratio for direct (above) and transitive (below) dependents. This ratio represents the additional impact of a package due to its presence in more than one ecosystem. 
The y-axis indicates the percentage of x-sets having at least the corresponding ratio value. 
We observe from Figure~\ref{fig:rq3direct} that 7.7\% of the x-sets (148 x-sets) have a global potential that is at least 1.2 times greater than the maximal local impact (i.e., ratio $>=1.2$). 
For transitive dependents (Figure~\ref{fig:rq3transitive}), 
only 5\% have a ratio greater than 1.33.
Thus, \textit{a small set of x-sets can impact a multitude of other packages in other ecosystems, but this impact can be significant given the number of dependent packages}; up to 54,751 direct and 332,504 transitive dependent packages.

\subsection*{$RQ_4$ Do cross-ecosystem packages have different characteristics compared to other packages in the ecosystem?}

With the fourth research question, we aim to investigate if the characteristics of cross-ecosystem packages are different than the characteristics of packages being distributed in a single ecosystem. 

The cross-ecosystem packages only represent a small fraction ($<1\%$) of the total number of packages in each ecosystem. As reported in detail in Section~\ref{sec:results}, only 15,389 out of 1,556,300 packages are cross ecosystem packages. To avoid any bias in our analyses when comparing the characteristics of cross-ecosystem packages with those of other packages, we generated representative samples of each ecosystem's package population that are comparable in size with the population of cross-ecosystem packages in each ecosystem. 

The research literature reports different approaches to sample a population of software projects. For example, the approach by Nagappan et al.~\cite{Nagappan:2013:DSE} aims to capture the diversity of values of the projects' characteristics so that the sample will maximize the coverage of the population. In our analysis, we aim to find, if any, the differences of cross-ecosystem packages w.r.t a representative sample of the ecosystem's package population.
The aforementioned algorithm would introduce bias in our analyses as we need to compare between cross-ecosystem packages and a representative sample of the population, and not one that covers as many projects in the population as possible.

\begin{algorithm}
  \algsetup{linenosize=\small}
  \small
  \caption{Random Sample Generation}
  \label{alg:random_gen}
  \begin{algorithmic}[1]
    \REQUIRE population P, attributes A, sample size S, stop criterion ST
    \STATE $sample \leftarrow \emptyset$
    \STATE $same\_attributes \leftarrow \emptyset$
    \STATE $repetitions=0$
    
    \WHILE{$True$}
    \STATE $repetitions=repetitions+1$
    \STATE $tsample \leftarrow random\_sample(P)$
    \STATE $sample\_same\_attributes \leftarrow \{attr \in A | distr(attr,tsample)==distr(attr,P)\}$ 
    
    \IF{$|$sample\_same\_attributes$|>|same\_attributes|$}
    \STATE $sample \leftarrow tsample$
    \STATE $repetitions=0$
    \STATE $same\_attributes \leftarrow sample\_same\_attributes$
    \ENDIF
    
    \IF{$|$sample\_same\_attributes$|==|A|$ OR $repetitions>ST$}
    \STATE $break$
    \ENDIF
    \ENDWHILE
    
    \STATE \RETURN $sample$
  \end{algorithmic}
\end{algorithm}

Therefore, we implemented a random sample generation (Algorithm~\ref{alg:random_gen}) that will preserve the characteristics of the population, i.e., if there is a predominant class in a characteristic for the entire population, we aim for this characteristic to be predominant in the random sample as well.
Algorithm~\ref{alg:random_gen} presents the steps to generate a random sample of size S, originating from a population P and based on a set of attributes A. The random sample is generated (line 6) and then, the set of attributes $sample\_same\_attributes$ is retrieved for which the random sample follows the same distribution as the entire population (line 7). If the set $sample\_same\_attributes$ is larger than the current best (line 8), then the random sample and the set of attributes are stored (lines 9 and 11). Each time the algorithm fetches a new random sample, the $repetitions$ counter increments (line 5); if the algorithm finds a better match then the counter is reset to zero (line 10). The algorithm finishes when all attributes of the random sample follow the same distribution as the population, or when the number of repetitions without finding a better sample exceeds a threshold $ST$ (line 13). 


For our analysis, we gathered a random sample for each ecosystem according to Algorithm~\ref{alg:random_gen}, where $P$ is the set of non cross-ecosystem packages, the sample size $S$ is the number of cross-ecosystem packages, the stop criterion $ST=100$, and the set of attributes $A$ contains the following information provided by libraries.io for each git repository: \# of versions, repository size, \# of stars, \# of forks, \# of open issues, \# of watchers, \# of contributors and longevity.
We compared the cross-ecosystem packages with the random sample generated for each ecosystem by use of Mann-Whitney U tests to determine which attributes are different between the two samples; all tests were performed with $\alpha = 0.05$. 
The resulting p-values were adjusted following Bonferroni-Holm method to control family-wise error rate.

%
%

\begin{table}[t]
\centering
  \caption{Comparison of 8 attributes $\in A$ between cross-ecosystem packages and randomly sampled packages. The arrows indicates if cross-ecosystem packages have larger ($\uparrow$) or lower values ($\downarrow$).}
  \scalebox{0.88}{%

    \begin{tabular}{lp{1.2cm}p{3.5cm}p{2.6cm}}
 \hline
    \textbf{ecosystem} & \textbf{\# of different attributes} &  \textbf{larger values  ($\uparrow$)} &  \textbf{smaller values  ($\downarrow$)} \\ 
    \hline
    CPAN &  0/8 & $\emptyset$ & $\emptyset$ \\
    CRAN   & 0/8 & $\emptyset$ & $\emptyset$ \\
    Cargo  & 1/8 & \{stars\}& \\
    Clojars & 8/8 & $A$ $\setminus \{\text{longevity}\}$ &  $\{\text{longevity}\}$  \\
    CocoaPods& 7/8 &  
    $A \setminus \{versions, longevity\}$  & 
    $ \{\text{longevity}\}$\\  
    Hackage &  0/8  & $\emptyset$ & $\emptyset$ \\
    Maven &  7/8 & $A \setminus \{versions,longevity\}$& 
    $\{\text{longevity}\}$\\
    npm   &  7/8 &  $A \setminus \{\text{longevity}\}$ & $\emptyset$\\
    NuGet &   8/8 & $A$  $\setminus \{\text{longevity}\}$  & 
    $\{\text{longevity}\}$\\
    Packagist& 8/8 & $A$ $\setminus \{\text{longevity}\}$& 
    $\{\text{longevity}\}$\\
    Pypi  &  7/8 & $A \setminus \{\text{longevity}\}$  & $\emptyset$
    \\
    Rubygems& 7/8  &  $A \setminus \{\text{longevity}\}$& $\emptyset$\\ \hline
    \end{tabular}%
    }
  \label{tab:rq4-diffs}%
\end{table}%

Table~\ref{tab:rq4-diffs} presents the different attributes for each ecosystem.
These results suggest that \textit{cross-ecosystem packages in CPAN, CRAN and Hackage do not have different characteristics than other packages}, while Cargo cross-ecosystem packages only have a larger \emph{\# of stars} compared to the randomly sampled ones. For the remaining ecosystems, cross-ecosystem packages are either completely different (Clojars, NuGet, Packagist), or differ in most attributes (CocoaPods, Maven, npm, Pypi, Rubygems). 
Overall, \textit{cross-ecosystem packages have usually more stars, forks, watchers and contributors w.r.t. other packages in the ecosystem} (see the listed attributes on the third and fourth columns of~Table~\ref{tab:rq4-diffs}). 
By taking into account that cross-ecosystem packages tend to be more popular based on stars, forks and watchers, and that they have a large number of dependent packages across different ecosystems (see $RQ_3$), it is evident that these packages are very important and more in-depth analyses are required to study their evolution.

\section{Conclusion}
\label{sec:conclusion}

We observed that a small fraction of the software packages distributed in packaging ecosystems are also distributed in packaging ecosystems for other programming languages. Such cross-ecosystem packages can benefit from a potentially larger pool of contributors. On the other hand, issues in these packages (e.g. bugs or security issues) may have a wider impact, because they may affect other packages in each ecosystem in which they are distributed.
In the preliminary investigation of this paper, we empirically analysed the characteristics of cross-ecosystem packages in 12 popular programming language ecosystems. We observed that many cross-ecosystem packages may have a potentially high impact in multiple ecosystems due to the large number of other packages depending on them.

More in-depth analyses are required to shed more light in the reasons for distributing packages across multiple ecosystems, as well as the evolution dynamics and release policy of these packages. We also aim to study if there are socio-technical health issues in the development teams of cross-ecosystem packages.
These and related aspects and questions will be investigated in our future work.




\section*{Acknowledgment}
This research was carried out in the context of FRQ-FNRS collaborative research project R.60.04.18.F ``SECOHealth'', FNRS Research Credit J.0023.16 ``Analysis of Software Project Survival'' and Excellence of Science project 30446992 \textsf{SECO-Assist} financed by FWO - Vlaanderen and F.R.S.-FNRS.

\newpage

\bibliographystyle{unsrt}
\bibliography{IEEEfull,references}

\end{document}